\documentclass[aps,prb,superscriptaddress,twocolumn]{revtex4-1}

\pdfoutput=1 

\usepackage{graphicx}
\usepackage{bm}
\usepackage{dcolumn}
\usepackage{amssymb}
\usepackage{amsmath}
\usepackage{amsfonts}
\usepackage{color}
\usepackage{epsfig}

\def \Z2{$\mathbb{Z}_2$}

\newcommand{\bra}[1]{\langle #1|}
\newcommand{\ket}[1]{|#1\rangle}

\newcommand{\be}{\begin{equation} }
\newcommand{\ee}{\end{equation} }
\newcommand{\ba}{\begin{eqnarray} }
\newcommand{\ea}{\end{eqnarray} }

\newcommand{\bpm}{\begin{pmatrix}}
\newcommand{\epm}{\end{pmatrix}}
\newcommand{\bmm}{\begin{matrix}}
\newcommand{\emm}{\end{matrix}}

\newcommand{\mtwo}[4]{\left( \begin{array}{c c} #1 & #2 \\
	#3 & #4 \end{array} \right)}

\begin{document}

\title{
    Interacting and fractional topological insulators via the $\mathbb{Z}_2$ chiral anomaly.}
\author{Maciej Koch-Janusz}
\affiliation{Department of Condensed Matter Physics, Weizmann Institute of Science, Rehovot, Israel.} %
\author{Zohar Ringel}
\affiliation{Department of Condensed Matter Physics, Weizmann Institute of Science, Rehovot, Israel.}
\affiliation{Theoretical Physics, Oxford University, 1, Keble Road, Oxford OX1 3NP, United Kingdom.}

\begin{abstract}
Recently it was shown that the topological properties of $2D$ and $3D$ topological insulators are captured by a $\mathbb{Z}_2$ chiral anomaly in the boundary field theory. It remained, however, unclear whether the anomaly survives electron-electron interactions. We show that this is indeed the case, thereby providing an alternative formalism for treating topological insulators in the interacting regime. We apply this formalism to fractional topological insulators (FTI) via projective/parton constructions and use it to test the robustness of all fractional topological insulators which can be described in this way. The stability criterion we develop is easy to check and based on the pairswitching behaviour of the noninteracting partons. In particular, we find that FTIs based on bosonic Laughlin states and the $M=0$ bosonic Read-Rezayi states are fragile and may have a completely gapped and non-degenerate edge spectrum in each topological sector. In contrast, the $\mathbb{Z}_{k}$ Read-Rezayi states with $M=1$ and odd $k$ and the bosonic $3D$ topological insulator with a $\pi/4$ fractional theta-term are topologically stable.  
\end{abstract}
\pacs{73.43.-f, 73.43.Cd, ...} %

\maketitle


\section{Introduction}

Topological insulators (TIs) have attracted a large amount of attention in recent years due to their novel bulk and surface properties\cite{Hasan2010}. The bulk of these materials is insulating and characterized by topological indices which measure certain twists in the band structure.  The topology of the bulk implies, via the bulk-edge correspondence, that the surfaces of these materials are necessarily metallic. 

The theories that emerge on the boundaries of topological phases can be understood as fractions of more standard ones\cite{Qi2008,Liu2012}. For example, the integer quantum Hall (IQHE) edges can be thought of as half of a Luttinger liquid, the quantum spin Hall effect (QSHE) as half of a spinfull wire and the strong topological insulator (STI) as half of the Bernevig-Zhang model at the transition. Of particular interest recently are the fractional topological insulators in 2D\cite{AdyLevin} and 3D\cite{Maciejko} (FTIs), which can host anyonic and even non-Abelian excitations\cite{MaciekLevinAdy}.

The question of stability of the edge theories of topological insulators in the nonintereacting case is, even in the presence of disorder, well understood. As long as the time-reversal (TRS) and charge conservation symmetries are not broken, the edge is stable.
When interactions are considered, an important distinction arises. For the 2D and 3D topological insulators which can be adiabatically connected to noninteracting band insulators various approaches indicate that their edge states are robust\cite{RyuMoore,AdyLevin,Wang2010,Gurarie2011}. This means that their boundaries cannot be completely gapped out -- they will either have a gapless boundary or a protected degeneracy associated with the boundary.

Topological insulators which cannot be adiabatically connected to band insulators present a more subtle challenge. A prominent class of examples of such phases are the FTIs, i.e. TRS analogs of fractional quantum Hall states. For the two dimensional FTIs supporting only Abelian excitations, which admit a K-matrix description, the $\mathbb{Z}_2$ classification has been shown to persist. Reference \onlinecite{AdyLevinStability} establishes an elegant general formula for the value of this index in terms of the ratio $\sigma_{sH}/e^*$ of quantum spin Hall conductivity and the value of the smallest possible electric charge, in appropriate units. The same criterion has been derived before in less general settings\cite{AdyLevin,Mudry1,Mudry2}. The case of non-Abelian systems or extension to three dimensions has not yet been settled\cite{Cappelli, Cappelli-misc}.

In a recent paper \cite{ZoharAdy2013} it was shown that the properties of noninteracting TIs can be described in a compact form using a novel field theory anomaly dubbed the "$\mathbb{Z}_2$ chiral anomaly". The basic topological properties of TIs such as pair switching, time-reversal-parity pumping and the $\mathbb{Z}_2$ topological index algebra\cite{FuKane2006} are manifested in the properties of field-theoretical quantities such as the partition and correlation functions. 

In this work we show that the $\mathbb{Z}_2$ chiral anomaly survives interactions and is therefore a fully robust feature of the field theory. We then use this anomaly as a diagnostic tool to determine the stability of various parton constructions of fractional TIs in $2D$ and in $3D$. A convenient aspect of the anomaly approach is that it works within a fermionic description of the edge theory and it is unaffected by the gauge field which glues the partons into physical entities. Consequently, analyzing the robustness of a candidate fractional TI amounts to analyzing the anomaly content of the free parton theory or equivalently their pairswitching behavior. Provided that the free theory is anomalous (or performs pairswitching) the phase is stable. Stability here should be understood in the following limited sense: a single low-lying excited state is guaranteed to exist on each boundary and within each topological sector. 

Our results are consistent with the possibility of spontaneous TRS breaking\cite{AdyYuval} and with other more exotic phase transitions, for instance to a gapped topological surface phase\cite{Metlitski,Fidkowski,Potter}. Considering the former, provided that no infinitesimal ordering field is introduced, the system will be in some superposition of the two TRS-related ground states. Since the symmetric and antisymmetric superpositions are exponentially close in energy, there is always an (exponentially) low-lying excitation above the ground state. If an infinitesimal TRS-breaking ordering field is introduced, our arguments no longer apply. In the case of transition to a gapped topological surface phase our results imply that there must be ground state degeneracy on a torus. Indeed for the phases discussed in Refs. \onlinecite{Metlitski,Fidkowski,Potter}, such a degeneracy is present. It is also worth noting that we do not rely on translational symmetry of the problem, hence the results are valid in the presence of disorder.

We apply the anomaly approach to various FTIs. In particular, we consider two dimensional FTIs made of a TRS pair of bosonic or fermonic Laughlin wave functions. The former are shown to be unstable and the latter stable. Fractional TIs made of $\mathbb{Z}_k$ Read-Rezayi states are stable only when they are fermionic (i.e. when $M$ is odd) with $k$ odd. The $3D$ bosonic topological insulator with a $\pi/4$ fractional theta term\cite{Senthil} is shown to be stable. 

This work is organized as follows: in section \ref{Sec-II} we recall some of the essential properties of the $\mathbb{Z}_2$ anomaly, its relation to pairswitching \cite{FuKane2006} and the existence of gapless edge excitations. This builds the technical background required in section \ref{Sec-int}, where we establish the stability of the anomaly to interactions. Section \ref{Sec-part} shows how the bulk and edge theories of FTIs can be described via parton or projective constructions and also how the anomaly approach applies to these theories. In section \ref{Sec-examples} we establish the robustness or fragility of a variety of FTIs based on the anomalous/pairswitching properties of their parton edge theory. We conclude with a discussion and outlook section (Sec. \ref{Sec-discussion}).

\section{The $\mathbb{Z}_2$ Chiral anomaly}
\label{Sec-II}
An anomaly in quantum field theory refers to the situation where a symmetry of the Lagrangian,
which is present at the classical level, is lost in quantization. A prototypical example is the chiral
anomaly of the quantum electrodynamics in 3+1 dimensions or the chiral anomaly in 1+1 dimensions describing the edge of a quantum Hall system \cite{peskin,DHLee1996}. The presence of the anomaly implies, among
others, that the response of the system to the applied external field will also not obey the symmetry
of the Lagrangian. In the quantum Hall case, where edge theory possesses a chiral anomaly \cite{DHLee1996}, the  anomalous current can be understood as coming from the Hall current in the bulk. 

This interplay between the bulk and the boundary is a general feature and the field theory formulations of boundaries of topological phases are usually associated with anomalies \cite{RyuMoore}. A rough intuition based on the previous examples is that the bulk allows conserved quantities to escape the surface thus providing a physical means to the symmetry violation. 

As shown in Ref. \onlinecite{ZoharAdy2013}, topological insulators exhibit an anomaly associated with TRS and charge conservation symmetry. The relevant field theory is simply that of two TI edges at the two ends of a long TI cylinder.  The main field theory quantity associated with this anomaly is the partition function itself, which vanishes following a flux insertion through the cylinder. This behavior is topologically robust and directly linked to the fact that the edge spectrum performs a pair switching behavior as a function of the flux. Its physical meaning is simply an orthogonality between the state before and after the flux insertion. To show that this is an anomaly in the more strict sense of a symmetry violation, one can show that a symmetry forbidden 2-point correlation
function involving creation and annihilation operators on two decoupled edges diverges in the limit
$m \longrightarrow 0$, where $m$ is the strength of the coupling between the edges (see also Eq. (\ref{green1}) below).

Let us now explain these results in more detail. The Euclidean action describing the two edges of the topological insulator on a cylinder considered in Ref. \onlinecite{ZoharAdy2013} is given by:
\begin{equation}\label{anom1} S = \int dxd\tau \bar{\psi}_{\sigma}[\hat{S}_{ch}]_{\sigma\sigma'}\psi_{\sigma'} ,\end{equation}
\begin{equation}\label{anom2} \hat{S}_{ch} = (i \hbar \partial_{\tau} +i\mu)\sigma_x + v_0\sigma_y\mathcal{H}_{edge}[A_x] \equiv \mtwo{0}{D}{D^{\dagger}}{0},\end{equation}
where $\mathcal{H}_{edge}[A_x]$ specifies the low-energy effective hamiltonian of the edge of the topological insulator, which also includes the time-dependent electromagnetic field. The variables $\bar{\psi}$ and $\psi$ are 4-component spinors given by $\bar{\psi} = (\bar{\psi}_d,\bar{\psi}_u)$ and $\psi = (\psi_u^T,\psi_d^T)^T$, where the components are themselves spinors in the spin space, the indices $u,d$ refer to upper/lower edge of the cylinder, or in the chiral language: $\bar{\psi} = (\bar{\psi}_+,\bar{\psi}_-)$ and $\psi = (\psi_+^T,\psi_-^T)^T$. The indices $\sigma, \sigma'$ are in the edge/chirality space. In this particular form the action includes the rotation of the $\bar{\psi}$ spinor by $i\sigma_x$. The "chirality" in the above should not be confused with the direction of propagation of electron on the edge.

Let us, for completeness, recall the definition of time reversal symmetry (TRS) in our system:
\begin{equation}\label{trs1} \mathcal{T}\hat{S}_{ch}\mathcal{T}^{-1} = (is_y\sigma_x) [\hat{S}_{ch}]^T (is_y\sigma_x)^T, \end{equation}
where the Pauli matrices $s$ act on the spin degree of freedom. Equivalently, the action of time-reversal on the spinor fields is given by:
\begin{gather}\nonumber \bar{\psi}(t) \longrightarrow \psi^T(-t) [s_y\sigma_x]^T \\
\label{tr2}  \psi(t) \longrightarrow s_y\sigma_x \bar{\psi}^T(-t). \end{gather}

One can now envisage the following thought experiment: let us put the topological insulator on an annulus and change the flux threading the hole of the annulus adiabatically from $-\Phi_0/2$ to $\Phi_0/2$. Denote by $\ket{gs}$ the groundstate at $-\Phi_0/2$ and by $G\ket{gs}$ the groundstate at $\Phi_0/2$. Let $U$ be the time-evolution operator which carries out the flux insertion at a rate $(\Delta T)^{-1}$. The flux insertion time, $\Delta T$, is taken to be larger than the inverse bulk gap. We are interested in the overlap of the state obtained by threading the flux with the groundstate, which is given by the following expression \cite{ZoharAdy}:
\begin{equation}\label{anom4} \bra{gs}G^{\dagger}U\ket{gs}  = Z[m] \equiv \int D[\bar{\psi}\psi]e^{-S_0(m)[\bar{\psi}\psi]}, \end{equation}
where the noninteracting action $S_0(m)$ contains additionally a time dependent gauge-field capturing the flux insertion ($A_x = \frac{h t}{\Delta T e L}$) and a source term $m\bar{\psi}[\sigma_0\otimes s_0]\psi$, which physically corresponds to coupling of charges on both edges.  For a trivial band insulator, this overlap is equal to 1, up to a phase. In contrast, it is 0 for the topological insulator. More precisely,  it vanishes as $m^2$, as $m$ goes to zero. Therefore, threading a flux will bring us to an orthogonal state in the case of the topological insulator. 

We note that the above equality is obtained for an action with a mixed Euclidean and real time contour, as described in Ref. \onlinecite{ZoharAdy}. The vanishing of the l.h.s also extends to pure Euclidean time by taking $A_x = \frac{h \tau}{\beta e L}$ where $\beta$ is the inverse temperature. For concreteness we shall focus on such Euclidean-time flux insertions, although this plays no essential role in rest of the paper. 

The vanishing of the partition function, and hence the overlap, is intimately linked to the appearance of action zero modes -- if we formally compute the expression Eq. (\ref{anom4}) we obtain:
\begin{equation}\label{anom6} \bra{gs}G^{\dagger}U\ket{gs} = Det[\hat{S}_{0}(m)] = \prod_n  \beta(\lambda_n + m), \end{equation}
whence it is easy to see that should some $k$ action eigenvalues $\lambda_n$ of the action Eq. (\ref{anom2}) be zero, the partition function will vanish as $m^k$.
Crucially, the existence of a pair of action zero modes for the topological insulator is guaranteed by the result obtained in Ref. \onlinecite{ZoharAdy2013}. It can be stated in the following way:
\begin{equation} \label{nu2eqPair} \nu_2 = Pairswitching, \end{equation}
where the new topological index $\nu_2$ is given by $ \nu_2 = DimKer[D] \ \ mod \ 2$. Intuitively, $\nu_2$ being equal to $0$ or $1$ corresponds to the presence or absence of a TRS protected pair of zero modes. The 'pairswitching' refers to the states changing their Kramers partner during a half-flux insertion.
This result is an extension of the Atiyah-Singer-Patodi theorem to the case of noninteracting topological insulators. 

Equivalently, one may consider a symmetry forbidden 2-point correlation function $G_{hop} = \langle \int dxdt\ \bar{\psi} \sigma_0 s_0 \psi \rangle$ involving creation and annihilation operators on two decoupled edges. Naive application of chiral symmetry suggests it should vanish, as applying the chiral transformation changes the function sign: 
\begin{equation}\label{anomgreen} \langle \bar{\psi}_+\psi_+ \rangle \longrightarrow \langle \bar{\psi}_+e^{i\pi/2}e^{i\pi/2}\psi_+ \rangle = -\langle\bar{\psi}_+ \psi_+\rangle, \end{equation} 
however a direct calculation produces the following result\cite{ZoharAdy2013}:
\begin{gather}\label{green1} G_{hop}  = \partial_m\log\left( \int d\bar{\psi}d\psi e^{-\int dxdt \bar{\psi}\hat{S}_{0}(m) \psi }  \right) = \\
\nonumber = \partial_m\log\left( \prod_n  \beta(\lambda_n + m) \right) = \frac{\nu_2}{m} + O(m^0,m^1,\ldots).  \end{gather}
Thus the anomaly is also associated with the $1/m$ pole in the Green's function (we take the limit $m\longrightarrow 0$ while keeping the system size and the temperature fixed).

Again, the $1/m$ pole has its origin in the action zero modes, as is evident from the denominators in the energy representation of the Green's function:
\begin{equation}\label{resolvent} \hat{G}(m) = \sum_n \frac{\ket{n}\bra{n}}{\lambda_n+m}. \end{equation}
Furthermore, considering the following expression involving the 'regular part', i.e. the part that does not contain zero-modes:
\begin{gather}\label{regular1} \bra{+}\sum_{n\not= 0} \frac{\ket{n}\bra{n}}{\lambda_n} \ket{+} =  \\ 
\nonumber = \bra{+}\sum_{n\not= 0} \frac{\ket{n}\bra{n}}{\lambda_n} + \sum_{n \not= 0} \frac{\ket{\sigma_z n}\bra{\sigma_z n}}{-\lambda_n}\ket{+} = 0, \end{gather}
we conlude that it vanishes between the states of like chirality, thus the 'regular' Green's function and hence also the nonzero-modes play no part in the anomaly. In the above calculation we used the fact that for each action eigenstate $\ket{n}$ with an eigenvalue $\lambda_n \not= 0$ there exists an action eigenstate $\sigma_z \ket{n}$ with the eigenvalue $-\lambda_n$, which is due to the chiral action Eq. (\ref{anom2}) anticommuting with $\sigma_z$. 

We would like to extend the above considerations by including interactions. The action zero-modes will be central to the further discussion, let us therefore introduce some notations. Let $\varphi_0$ and $\varphi_{\bar{0}}$ denote the action zero modes with $\sigma_z = \pm$, respectively, i.e. the solutions to the equations  $D\varphi_0 = 0$ or $D^{\dagger}\varphi_{\bar{0}} = 0$. We shall define the \emph{zero-mode components} of the field $\psi$ in the following way:
\begin{gather}\label{zmc}\nonumber \psi_0 = \int dxd\tau \varphi_0^*(x,\tau)\psi_+(x,\tau) \\
\psi_{\bar{0}} = \int dxd\tau \varphi_{\bar{0}}^*(x,\tau)\psi_-(x,\tau) \end{gather}

\begin{figure}[tb]
\centerline{
\includegraphics[trim = 60mm 40mm 10mm 40mm,width=1.0\columnwidth]{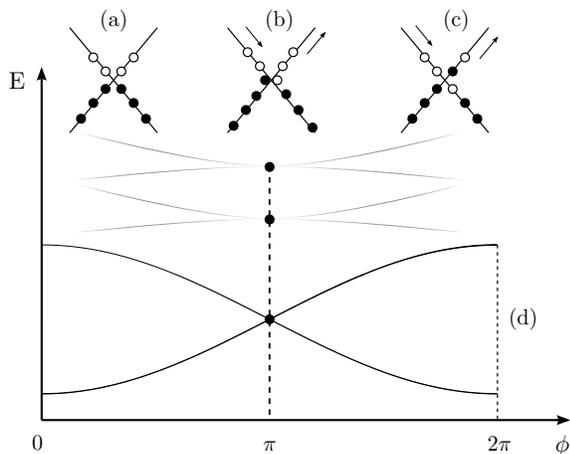}
}
\caption{Spectral motions of the non-interacting low-energy many body spectrum and occupation of single-particle momentum states as a function of the adiabatically threaded flux $\phi$. Insertion of half of a flux quantum results in one electron being pumped to the edge and a Kramer's degeneracy corresponding to momentum state occupation shown in Fig. \ref{anomfig1}b. Completing the flux insertion one arrives at an orthogonal state with momentum-state occupation as in Fig. \ref{anomfig1}c. Alternatively, the many-body groundstate of a TI evolves into an orthogonal state following a full-flux insertion and exhibits a level crossing at $\pi$ flux as shown in Fig. \ref{anomfig1}d.}
\label{anomfig1}
\end{figure} 

\section{Robustness of the $\mathbb{Z}_2$ chiral anomaly to interactions}
\label{Sec-int}
In this section we show that the $\mathbb{Z}_2$ anomaly is robust to electron-electron interactions provided that (i) the bulk gap remains open and (ii) there is no direct coupling between the gapless boundaries. Specifically, the orthogonality relation implied by Eqs. (\ref{anom6},\ref{nu2eqPair}) holds also for interacting TIs. Another way to phrase it is to say that the boundary of an interacting TI always supports a low-lying state which can be excited by a full-flux insertion. 
In the next sections, we will generalize this results to fractional TIs. 

In the adiabatic limit, defined here as the limit in which the rate of the flux insertion is much smaller than the level spacing on the boundary, this result can be understood by arguments similar to those used in Refs. \onlinecite{FuKane2006,AdyLevin}. Consider the many-body spectrum of a single, translation-invariant edge of a non-interacting TI with a fixed particle number. For simplicity, let us assume that prior to the flux insertion there is no Kramer's degeneracy and the occupation of each single-particle momentum state is that appearing in Fig.\ref{anomfig1}a, where full circles denote occupied states. Upon inserting half of a flux quantum, one electron is pumped to the edge resulting in a Kramer's degeneracy and a momentum state occupation  shown in Fig.(\ref{anomfig1}b). Completing the flux insertion one arrives at an orthogonal state with momentum-state occupation as in Fig.(\ref{anomfig1}c). The low-lying spectrum of the noninteracting many-body system, depicted in Fig.\ref{anomfig1}d as a function of flux, exhibits a level crossing at $\pi$ flux. 

The presence of a crossing in the the spectrum as a function of flux is a robust feature of the many-body spectrum, even in the presence of interactions. Indeed, provided that TRS is not \emph{explicitly} broken and that edges remain decoupled, the Karmer's degeneracy point at $\pi$ flux cannot be removed. Thus, provided that the flux insertion is carried at a rate much slower than the splitting at $\phi=0$, a level crossing occurs and the ground state evolves into an excited state. Notably, this argument holds also in the presence of a spontaneous TRS breaking on the edge -- in this case the two states which cross will be the two symmetry-related ground states. The splitting between, and hence also the rate of flux insertion, will be exponentially small in the system size. In fact, the only way to remove this behaviour without explicitly breaking TRS is to couple the two boundaries, for example via a gapless excitation in the bulk, thus allowing the spectral motions of the different boundaries to unwind together. 

Next we re-establish this result using the anomaly approach. This is done by showing that the partition function appearing in Eq. (\ref{anom4}) remains zero also when the action is non-gaussian and contains interactions.  This field theoretical approach has some advantages over the above simpler considerations. First, in the presence of fluctuating gauge fields, it can be used to address the orthogonality relations in a gauge invariant way. This will be useful when we turn to consider projective/parton constructions of fractional TIs, where such gauge fields naturally emerge. Furthermore the anomaly formalism will allow us to show that the above orthogonality result persists even for non-adiabatic flux insertions.   

To this end, we first choose a convenient regulator which is a sharp cut-off on the non-interacting action spectrum. This cut-off has been used before to analyze the chiral anomaly\cite{Fujikawa}. Given a finite system in a finite temperature the path integral representation of the partition function reduces to a finite number of integrals.
The interactions we consider are generic density-density $\bar{\psi}\sigma_{\pm}s_0\psi$ or spin-spin $\bar{\psi}\sigma_{\pm}s_x\psi$ type terms on one or the other edge, which we add to the action:
\begin{equation}\label{anom9} S_{int} = g_0^{\pm}\left(\bar{\psi}[\sigma_{\pm}\otimes s_0]\psi\right)^2 + g_x^{\pm}\left(\bar{\psi}[\sigma_{\pm}\otimes s_x]\psi\right)^2. \end{equation}
It is easy to verify that a term like $\bar{\psi}\sigma_{+}s_0\psi$ acts on a single edge, but couples different chiralities.

We next use the idea of integrating out high energy degrees of freedom to derive an effective low energy action. We take this idea, however, to its absolute extreme, that is, we integrate out everything but the zero-mode components. 
\begin{gather}\nonumber Z_{int}[m] = \int d[\psi]  e^{-S_0(m) + S_{int}} = \\
\nonumber \int \prod_{\alpha=0,\bar{0}}d[\psi_\alpha]\int\prod_{n\not= 0} d[\psi_n] e^{-S_0[\psi_0] -S_0[\psi_n] + S_{int}[\psi_0,\psi_n]} = \\
\label{rgstyle1}= \int d[\psi_0]d[\psi_{\bar{0}}] \ e^{-S_0[\psi_0] + S_{eff}[\psi_0]}, \end{gather}
where $S_0[\psi_0]\equiv S_0[\psi_0](m) = im\bar{\psi}_0\psi_0 + im\bar{\psi}_{\bar{0}}\psi_{\bar{0}}$. The effective action obtained upon integrating out higher mode components is given by:
\begin{equation}\label{rgstyle2} S_{eff}[\psi_0] = \log\langle e^{-S_{int}[\psi_0,\psi_n]} \rangle_{n\not= 0}, \end{equation}
where $\langle\cdot \rangle_{n\not= 0}$ denotes the average over nonzero-mode components. We also used a compacted notation for the Grassmann integration over components: $d[\psi] \equiv d\bar{\psi}d\psi$.

The resulting effective theory is phrased in terms of the remaining four Grassmann variables which are the $\psi_0,\psi_{\bar{0}}$ and their conjugates. We wish to analyze what type of terms can be generated in this effective action based on symmetry considerations.  An essential property of the integrating out procedure is that it does not involve the zero mode components which are the only modes which actually violate the symmetry\cite{Fujikawa,Nakahara}. Thus the terms generated by this procedure must respect the classical properties of the action.  Most notably the fact that the edges are classically decoupled. For example, notice that the Green's function in the limit $m\longrightarrow0$ only breaks the chiral symmetry due to the presence of the zero modes (see discussion around Eq. (\ref{regular1})). In the diagrammatic language this integrating out procedure amounts to connecting the interaction vertices with 'regular' Green's functions, obtaining a set of new effective vertices for the zero mode components only. In what follows we work in the strict $m= 0$ limit.

We begin by rewriting the effective action as a cumulant expansion, using the following identity:
\begin{equation}\label{rgstyle3} \log\langle e^{-S_{int}[\psi_0,\psi_n]} \rangle_{n\not= 0} = \sum_{k=1}^{\infty} \frac{(-1)^k}{k!} \langle S_{int}^k \rangle_c[\psi_0], \end{equation}
where $\langle  \cdot\rangle_c$ denotes a cumulant. Thus we have:
\begin{gather}\nonumber S_{eff}[\psi_0] = -\langle S_{int} \rangle_c[\psi_0] +\frac{1}{2}\langle S_{int}^2 \rangle_c[\psi_0] - \ldots = \\ \label{rgstyle4} = -\langle S_{int} \rangle[\psi_0] + \frac{1}{2}\left( \langle S_{int}^2 \rangle[\psi_0] - \langle S_{int} \rangle^2[\psi_0]\right) - \ldots,\end{gather}
where the averages with respect to the non-zero mode components are given by:
\begin{gather}\label{rgstyle5} \langle S_{int}\rangle[\psi_0] \equiv \langle S_{int}[\psi_0,\psi_n]\rangle_{n\not= 0} = \\
\nonumber =  \frac{1}{Z_{n\not= 0}}\int d[\psi_n]\ S_{int}[\psi_0,\psi_n] e^{-S_0[\psi_n]}, \end{gather}
with $S_0[\psi_n]$ quadratic in the nonzero-mode components.

Next, we want to show that all the averages, and hence also the cumulants and the effective action vanish. To this end 
it is convenient to rewrite the term $S_{int}[\psi_0,\psi_n]$, which originates from expansion of the bare interaction term $S_{int} \propto \left( \bar{\psi}_{+}\psi_- \right)^2 + \left( \bar{\psi}_{-}\psi_+ \right)^2$ in mode components, as a product of zero-mode and nonzero-mode factors. The mode expansion of the bare interaction contains quartic terms, however its chiral structure does not allow more than two zero-mode components to appear in any term. This is because the $\psi_0, \psi_{\bar{0}}$ modes have a fixed chirality, hence expanding for instance $(\psi_-)^2$ in mode components, the only double zero-mode contribution that could appear is $\psi_0\psi_0$, which is clearly zero. In fact, the only allowed bare terms containing two zero-mode components are of the form $\bar{\psi}_{\bar{0}}\psi_{0}\bar{\psi}_{k}\psi_l$ or the hermitian conjugate thereof. There are no such restrictions on bare terms containing a single zero mode. 

Notice also that, due to the form of the interaction term $S_{int}$, we can -- as a bookkeeping device -- unambigously assign edge labels $u/d$ to \emph{single} zero-mode components and the corresponding nonzero-mode factors: for example the $\psi_0$ mode of positive chirality could only have originated from the $(\bar{\psi}_-\psi_+)^2 \propto (\bar{\psi}_u\psi_u)^2$ interaction term, hence we denote it by $\psi_0^u$. Since the bare interaction does not couple edges, the corresponding nonzero-mode factors have to carry the same edge label.

We can thus write the terms in $S_{int}[\psi_0,\psi_n]$ in the following form:
\begin{gather}\label{rgstyle6} S_{int}[\psi_0,\psi_n] = \hat{O}_0^u+ \hat{O}_0^d+ \psi_0^u\cdot\hat{O}_{\psi_0}^u + \psi_{\bar{0}}^d\cdot\hat{O}_{\psi_{\bar{0}}}^d +  \\
\nonumber \bar{\psi}_{\bar{0}}^u\psi_0^u\cdot\hat{O}_{\bar{\psi}_{\bar{0}}\psi_0}^u  + h.c., \end{gather}
where $\hat{O}$ contain only the nonzero-mode components. In particular, the $\hat{O}_0^{u/d}$ operator contains only nonzero-mode components which act exclusively on the upper or lower edge of the cylinder.

Since the averaging in $\langle S_{int} \rangle$ is done only over nonzero-mode components we have:
\begin{equation}\label{rgstyle7} \langle S_{int} \rangle[\psi_0]  = \psi_0^u\cdot\langle\hat{O}_{\psi_0}^u\rangle + \bar{\psi}_{\bar{0}}^d\cdot\langle\hat{O}_{\bar{\psi}_{\bar{0}}}^d\rangle + \ldots, \end{equation}
and the same factoring into zero-mode components and averages of nonzero modes holds for higher cumulants of $S_{int}$. Furthermore, the averages of higher powers of $S_{int}$ factor into products of averages on the upper and lower edges, since the action $S_0[\psi_n]$ we average it with does not couple the edges.

Using the above we now can show that the effective action vanishes order by order in zero-mode components. The terms in the new effective action cannot be of order higher than four in zero-mode components, since there are only four Grassmann variables left in the theory. Terms with an odd number of zero-mode components cannot be generated, since their coefficients are sums of averages of an odd number of nonzero-modes, which vanish due to charge conservation. 

A priori it is possible to generate three  distinct quadratic terms: \begin{gather}\label{quadterms1}\bar{\psi}_0^d\psi_0^u\cdot \langle \hat{O}_{\bar{\psi}_0}^d\hat{O}_{\psi_0}^u\hat{O}^{\Sigma}_0 \rangle \\
\label{quadterms2}\bar{\psi}_{\bar{0}}^u\psi_{\bar{0}}^d\cdot \langle \hat{O}_{\bar{\psi}_{\bar{0}}}^u\hat{O}_{\psi_{\bar{0}}}^d\hat{O}^{\Sigma}_0\rangle\\
\label{quadterms3} \bar{\psi}_0^d\psi_{\bar{0}}^d\cdot\langle \hat{O}_{\bar{\psi}_0}^d\hat{O}_{\psi_{\bar{0}}}^d\hat{O}^{\Sigma}_0 \rangle + \bar{\psi}_0^d\psi_{\bar{0}}^d\cdot\langle \hat{O}_{\bar{\psi}_0\psi_{\bar{0}}}^d\hat{O}^{\Sigma}_0\rangle + \ h.c., \end{gather}
where  we have defined $\hat{O}^{\Sigma}_0$ as:
\begin{equation}\label{quadterms2b} \hat{O}^{\Sigma}_0 = \pm \sum_{\alpha,\beta} \left( \hat{O}_0^u \right)^{\alpha}\left( \hat{O}_0^d \right)^{\beta}, \end{equation} 
and where the $\pm1$ ambiguity results from commuting all the $\hat{O}_0$ to the right.  

Notice however, that $\bar{\psi}_0^d\psi_0^u$ and $\bar{\psi}_{\bar{0}}^u\psi_{\bar{0}}^d$ explicitly break chiral symmetry, hence in order for the effective action to preserve the symmetry the averages in Eqs. (\ref{quadterms1},\ref{quadterms2}) multiplying them have to break it as well. This, of course, cannot happen, since these are regular averages which preserve classical symmetries of the action, as explained earlier. Thus the coefficient multiplying $\bar{\psi}_0\psi_0$ is necessarily zero. Analogous reasoning can be applied to the term $\bar{\psi}_0\psi_{\bar{0}}$, since $\bar{\psi}_0\psi_{\bar{0}}$ breaks time-reversal symmetry (the zero-mode components $0$ and $\bar{0}$ come with opposite spin hence this term is equivalent to a spin flip). Thus quadratic terms do not appear. The only remaining possibility is a quartic term $\bar{\psi}_0^d\bar{\psi}_{\bar{0}}^u\psi_{\bar{0}}^d\psi_0^u$.

We argue that this quartic term cannot appear either. Since it necessarily contains zero-mode components living on both upper and lower edges, the same is true of its coefficient, namely it has to contain averages of $\hat{O}$ operators from \emph{both} edges. Recall further that such averages factor into products of averages on both edges separately. This however leads to a contradiction, since the effective action is expressed in terms of the cumulants of $S_{int}$, which by definition correspond to fully connected diagrams or in other words, averages which \emph{do not} factor. Thus we arrive at the conclusion that in the effective action a quartic term cannot be generated, as its coefficient must factor. Hence its equal to zero. The same reasoning could have been applied to the chiral-symmetry breaking quadratic terms. 
We also note here that the diagrammatic approach offers another natural way to arrive at this conclusion: it is simply impossible to draw an effective quartic vertex starting from the interaction vertices we considered and connecting them with regular Green's functions.  

Thus far we have shown that in the strict $m=0$ limit for a finite system the partition function vanishes. We state here without proof that our reasoning generalizes to the case of finite $m$. Doing so we find that the partition function still vanishes as $m^2$. 

Lastly, we argue that these results are valid even if the edge undergoes a phase transition. This is not unreasonable since our proof is based on renormalization group reasoning rather than simple perturbation theory. This is not obvious either since our proof relies on an integrating-out procedure which is well defined for a finite system, however, for an infinite system it could perhaps generate infinite terms. To formally circumvent this subtlety we work with a finite system, calculate the anomaly and take the thermodynamic limit at the very end. Since the anomaly holds for every large but finite system size, it should also hold, by continuity, for an infinite system. We note in passing that scenarios in which TRS is broken spontaneously are perfectly consistent with this reasoning: as long as no infinitesimal ordering-field is added, the anomaly survives. It reflects the fact that the symmetric combination of the two TRS-equivalent ground states develops into the antisymmetric one following a flux insertion, as was explained in the discussion of Kramer's degeneracy in the beginning of this section. If an ordering field is introduced, TRS is broken explicitly and our proof is no longer valid. 

\section{Parton constructions of fractional topological insulators and the $\mathbb{Z}_2$ chiral anomaly}
\label{Sec-part}

Here we show how to extend the anomaly formalism to fractional topological insulators (FTIs) in order to analyze their stability. We do so using the parton\cite{Jain1989,WenParton2} and projective constructions\cite{WenParton1,Wen9}. Such constructions have been used to describe a variety of FTIs in both two\cite{Maciejko3} and three\cite{Maciejko2,Senthil,Ye} dimensions.  An FTI with an anomaly is guaranteed to have a low-lying boundary excitation within each topological sector. Conversely, for an FTI without an anomaly, at least in all of the examples we have considered, such an excitation can become gapped. 

The projective constructions are an elegant way of describing certain fractional quantum Hall states. They build upon a simple observation, that if one takes a product state of three species of electrons in a $\nu=1$ QH state, whose coordinates are given by the triple $z_i,w_i,x_i$ and imposes the constraint $z_i = w_i = x_i$, then the resulting wavefunction is that of a $\nu=1/3$ Laughlin state\cite{Jain1989}. Equivalently, in the field theory language, one rewrites the electron operator as a product of three fictitious partons and imposes the constraint by introducing an auxiliary $SU(3)$ gauge field to glue them. 
Remarkably, this conceptually simple construction accounts for a large number of non-Abelian states\cite{WenParton2,Wen9,WenParton1}.  

For example, the effective theory of a $\nu=1/N$ Laughlin state is described by the following bulk parton field theory\cite{WenParton2}: 
\begin{gather}
{\rm L} = \int d^2r dt \bar{\psi}_{\alpha} \left(i\partial_{\mu} - \frac{1}{N}A_{\mu} + a^{a}_{\mu}\tau^{a}_{\alpha \beta} \right)^2 \psi_{\beta} + \frac{1}{g} f^{a}_{\mu \nu} f^{a}_{\mu \nu}, \\
f^{a}_{\mu \nu} = \partial_{\mu} a^{a}_{\nu} - \partial_{\nu} a^{a}_{\mu} + f_{a b c } a^{b}_{\mu} a^{c}_{\nu}, 
\end{gather}
where $\mu,\nu = \{ x,y,t\}$, $A_{\mu}$ is the external electromagnetic gauge field, $a^{a}_{\mu}$ is the $SU(N)$ gauge field, $\tau^{a}$'s are the generators of $SU(N)$ in the fundamental representation, $f_{abc}$ is the structure factor of the Lie group, $g$ is the coupling strength and summation over repeated indices is implicit. The physical limit of the above theory is that of infinite coupling ($g \rightarrow \infty$). In this limit, the integration over the gauge field forces the $SU(N)$ currents to be locally zero thereby gluing the partons together to form physical electrons. The long wave length properties of the physical state can, however, be captured even in the weak coupling limit. Indeed, taking into account only the leading order fluctuations of the gauge field yields the right quasiparticle statistics and ground state degeneracy\cite{WenParton2}. 

Since the gauge coupling can be considered small, it is reasonable that the parton construction can be carried out directly on the low-energy theory of the edge. Such an approach was employed in Ref. \onlinecite{WenParton2} by dividing the free parton edge currents into the physical ones and the ones associated with $SU(N)$ gauge symmetry. The projection onto the physical Hilbert space was performed by removing the gauge currents. We are not aware of any parton construction in which this low-energy approach fails. 

In the field theory language, removing the unphysical currents amounts to taking the infinite coupling limit. However the low-energy features of the edge should be captured just as well with a strong yet finite coupling strength. This will make fluctuations away from the physical Hilbert space highly energetic thereby effectively excluding them from the low energy theory. Hence we write the low-energy theory of the edge as: 
\begin{gather}
{\rm L} = \int dt dx \bar{\psi}_{\alpha}  \left( i\partial_{\mu} + \frac{1}{N} A_{\mu} + a^{a}_{\mu}\tau^{a}_{\alpha \beta} \right)  \psi_{\beta}+ \frac{1}{g} f^a_{\mu \nu} f^a_{\mu \nu},
\end{gather}
with the coupling $g$ large but finite. 

We now turn to FTIs by taking two TRS-conjugate copies of the above parton edge theory. To analyze the anomaly content of this theory we again place it on a cylinder and perform the chiral transformation. The resulting theory for the edges of a FTI constructed from two Laughlin 1/N states is given by:  
\begin{gather}\label{fracanom1} S = \int dxd\tau\ \bar{\psi}_{\sigma}^\alpha[\hat{S}_{ch}]_{\sigma\sigma'}^{\alpha\beta}\psi_{\sigma'}^\beta + \frac{1}{g} f^a_{xt} f^a_{xt}, \\ \nonumber\hat{S}_{ch} = (i\hbar\partial_{\tau} - a^i_{\tau}\tau^i - e^*A_{\tau})\sigma_x + \sigma_ys_z(i\partial_{x} - a^i_{x}\tau^i - e^*A_{x}),\end{gather}
where $\sigma_i,s_i$ are the Pauli matrices in the chirality and spin space, $\tau^i$ are generators of $SU(N)$ and $\alpha,\beta$ are indices in the fundamental representation of the gauge group. The parton electric charge measured in units of $e$ is given by $e^*=1/N$. 

The question we address here is whether or not the above theory supports a low-lying excitation in the presence of TRS-respecting spin-mixing perturbations. Repeating the arguments given in the integer case, this can be analyzed by studying the partition function of this theory ($Z_{parton}$) in the presence of flux insertions. A zero partition function will imply an excited state, whereas a finite partition function will imply that the the ground state comes back to itself after an adiabatic flux insertion. 

Gauge fluctuations are clearly an important part of every parton theory. While certain quantities, such as the Hall conductance could be obtained correctly without including these fluctuations, properties such as the central charge of the edge conformal field theory do require fluctuations\cite{WenParton1}.  The $\mathbb{Z}_2$ anomaly is closer in spirit to the Hall conductance than to the central charge or the quasiparticle statistics. Indeed, it is a generalization of the chiral anomaly, the latter being the edge manifestation of the Hall conductance\cite{DHLee1996}. As we now argue, it is insensitive to gauge fluctuations. 

To establish this formally, we wish to transform the gauge fluctuations into parton-parton interactions and then apply the results of the previous section. To this end, we need to integrate out the gauge field so as to generate these effective interactions. Physically this procedure bears no meaning, as effective interactions will bind the partons together just as the gauge field did, it is simply useful to us from a technical point of view. The only obstacle here is gauge invariance, which implies that any integration over the gauge field, even over a local region, will formally be infinite. To remedy this we follow Ref. \onlinecite{peskin} and perform a gauge-fixing procedure using the Faddeev-Popov ghost system. This results in an action which contains extra massless fermionic ghost fields and a gauge symmetry violating term for the gauge field (introduced by the gauge-fixing procedure). Having removed the local gauge freedom, we can formally integrate out the gauge and ghost degrees of freedom to obtain an effective interaction for the partons. 

The resulting interacting parton system can be handled using a similar procedure to that used in Section \ref{Sec-int}. As argued previously, phase transitions within the edge theory do not affect the $\mathbb{Z}_2$ anomaly. Consequently, the fact that the free parton theory has a different central charge from that of the interacting partons does not play any role here. In essence, what matters is only the chiral structure, and this is unchanged. Loosely speaking, this situation is akin to QCD, where arguments based on the chiral anomaly of the free quarks allow for an accurate prediction of the neutral pion decay well below the confinement energy scale\cite{Adler1969}.  

We thus turn to discuss the free parton theory and its anomaly content. To this end we introduce a full-flux insertion in the form of a time-depended background $U(1)$ gauge field. The definition of a full-flux insertion changes with the parton theory. Without loss of generality we consider linearly time-dependent gauge field and put $A_{x}(\tau) = \frac{h}{\beta L e^*}\tau$, where $\tau$ runs from $0$ to $\beta$ and $e^*$ is the minimal charge in the system. This definition is equivalent to the existence of an unitary transformation $G$ which maps $H_{edge}(\tau=\beta)$ back to $H_{edge}(\tau=0)$. Note that the definition of a full-flux quantum depends on $e^{*}$, so that for instance in the case of the $\nu= 1/3$ Laughlin state threading three elementary flux quanta amounts to a single full-flux insertion. Technically speaking, the importance of demanding a full-flux insertion lies in the ability to define the boundary conditions for the action in a non-singular way\cite{ZoharAdy}. Physically speaking, this means that we are looking for an excitation within the same topological sector of the bulk. 

The parton theory is considered anomalous if a full-flux insertion, as described above, generates an odd number of action zero-mode pairs. Alternatively, one can check whether the joint spectrum of all the partons performs a pair-switching at half of a full-flux quantum. In contrast, the theory is considered trivial if there exist symmetry-respecting operators which remove all action zero-modes. In the next section we analyze the nature of various parton theories. Interestingly, for all the theories which we next analyze it holds that a non-anomalous theory is also trivial. This seems to hint at the possibility that the $\mathbb{Z}_2$ anomaly might be both a sufficient and a necessary criterion for the robustness of a FTI.   

\section{Examples}
\label{Sec-examples}

Here we analyze the stability of some candidate fractional TI phases, for which a projective construction is known. 

\subsection{Fermionic $SU(2N+1)$-based FTIs}

The simplest example is that of a FTI based on two time-reversal symmetric copies of a fermionic $\nu = 1/(2N+1)$ Laughlin state. We introduce $2N+1$ charge $e/(2N+1)$ partons $\psi_i$ per copy and write the electron operator $\Psi$ in the following way:
\begin{equation}\label{example1}
\Psi = \epsilon_{i_1i_2\ldots i_{2N+1}}\psi_{i_1} \psi_{i_2} ... \psi_{i_{2N+1}},
\end{equation}
where $\epsilon_{i_1i_2\ldots i_{2N+1}}$ is the Levi-Civita tensor. Since this tensor is invariant under $SU(2N+1)$ action we have a $SU(2N+1)$ gauge symmetry with matrices in the fundamental representation acting on the space spanned by the vectors $\psi_i$. 

This state is stable by the following argument: since the partons carry a fractional electric charge of $e^* = e/(2N+1)$, the full-flux insertion amounts to threading $2N+1$ flux quanta. The noninteracting action for a single spin species appears thus  as $2N+1$ copies of an IQHE action in the presence of a single flux quantum insertion. As discussed previously, $2N+1$ action zero modes will be generated per spin and therefore the $\mathbb{Z}_2$ topological index will be non-trivial: $\nu_2 = 1$.  Consequently the system is topologically stable.

\subsection{Bosonic $SU(2N)$-based FTIs}

We further consider a construction of a FTI based on two copies of a  bosonic Laughlin state\cite{Thomale-misc}. Here the gauge field is $SU(2N)$ and there are $2N$ spinfull partons carrying a fractional electric charge of $e^* = 1/2N$ each. In this case the previous argument involving a full-flux insertion  of $2N$ flux quanta suggests the system is unstable, as it creates an even number of zero modes, which, a priori, can gap each other out.

This is, however, not enough, as one should also verify that no other symmetry prevents pairs of zero modes from gapping each other out. In the field theory language this amounts to showing that the effective action for the zero modes allows for terms which respect the gauge, chiral and TRS symmetries. Considering for simplicity the $SU(2)$ case, we denote by $\psi_{1},\psi_{\bar{1}},\psi_{2},\psi_{\bar{2}}$ the zero-mode components associated with the two copies of the parton system. We find that the following two terms are fully consistent with all the symmetries of the action:
\begin{equation}\label{su2}
\bar{\psi}_{1}\bar{\psi}_{2}\psi_{\bar{1}}\psi_{\bar{2}} + \bar{\psi}_{\bar{1}} \bar{\psi}_{\bar{2}}\psi_{1}\psi_{2},
\end{equation}
where the first term is associated with the top edge and the second with the bottom edge of the cylinder. 
Given these terms the partition function is nonzero, as can be verified explicitly. To show this it suffices to notice that a term in perturbation theory generated by dropping a single power of either of the allowed terms in Eq. (\ref{su2}) will contain exactly one instance of each zero-mode component and hence will survive the Grassmann integration. Thus the system is not stable and there is no protected excitation within the same topological sector. 

One can further ask what the microscopic term which generates the above zero-more interaction term is. The form of the terms in Eq. (\ref{su2}) suggests that the appropriate perturbations, each acting on a single edge, should be proportional to $\Psi_{\uparrow}^\dagger \Psi_{\downarrow}$ and $\Psi_{\downarrow}^\dagger \Psi_{\uparrow}$, where the boson operator of each spin species (denoted by $\alpha = \uparrow/\downarrow$) is given by $\Psi_{\alpha} = \psi_{\alpha,1}\psi_{\alpha,2}$ in the $SU(2)$ case. Let us provide an intuitive argument for why this perturbation indeed removes the anomaly (the argument generalizes to the case of FTIs based on bosonic Read-Rezayi states, which we analyze next). To do that we focus on the system in the vicinity of the crossing point. We assume for simplicity that the chemical potential is exactly at Dirac point. We then study the level crossing at half of a flux quantum, where the occupation of single-particle states looks like two copies of the one shown in Fig. \ref{anomfig1}b, since each boson is made out of two equivalent partons. In the parton language our perturbation has the following form:
\begin{equation}\label{su22} \Psi_{\uparrow}^\dagger \Psi_{\downarrow} + h.c. = \psi^{\dagger}_{\uparrow,1}\psi^{\dagger}_{\uparrow,2}\psi_{\downarrow,1}\psi_{\downarrow,2} + \psi^{\dagger}_{\downarrow,1}\psi^{\dagger}_{\downarrow,2}\psi_{\uparrow,1}\psi_{\uparrow,2}. \end{equation}

The Hilbert space of the degenerate states at the crossing point can be described in terms of four fermionic states, which can be either filled or empty, with the restriction that exactly half of them are filled.  For intuition we again refer the reader to Fig. (\ref{anomfig1}b). We label them by $\ket{n_{\uparrow,1}n_{\downarrow,1}n_{\uparrow,2}n_{\downarrow,2}}$ with the occupation numbers $n_{\uparrow/\downarrow,1/2} = 0,1$. We now examine the action of the perturbation in Eq. (\ref{su22}) on this restricted space (first order degenerate perturbation theory). It is easy to see that the perturbation annihilates all but two eigenstates, which are given by $\ket{0101}\pm\ket{1010}$ and have eigenvalues $\pm 1$. These states are gauge-invariant and time-reversal symmetric. Depending on the sign of the perturbation, one of them is the unique lowest energy state. The perturbation therefore gaps out the crossing at half of a flux quantum. This implies that after an adiabatic full-flux insertion the ground state of the free partons with this perturbation returns to itself, rather than going to an orthogonal state. Since the perturbation is gauge invariant, it is reasonable that this result persists in the presence of gauge fluctuations.  

\subsection{Bosonic $M=0,\ \mathbb{Z}_k$ Read-Rezayi-based FTIs}
We also consider a parton construction based on two copies of $M=0$ $\mathbb{Z}_k$ Read-Rezayi states. Following Ref. \onlinecite{Wen9}, we introduce for each spin species a vector of $2k$ partons $\psi = (\psi_1,\psi_2, ... ,\psi_{2k})$ with charge $e^* = e/2$ each, and write the electron operator $\Psi$ as: 
\begin{equation}\label{zkrr}
\Psi = \psi^T \Lambda \psi, \mbox{\ \ \ \ }
\Lambda_{ij} = \sum_n [\delta_{2n,2n+1} - \delta_{2n+1,2n} ].
\end{equation}
This representation implies a $Sp(2k)$ gauge symmetry, with the $Sp(2k)$ matrices in the fundamental representation acting on the $\psi$ vector.

Since $e^* = 1/2$ or equivalently, since the Hall conductance is half of an integer, a double flux insertion is the minimal insertion which returns the system to the same topological sector. There is an even number of zero modes per spin which implies $\nu_2 = 0$ and thus there appears to be no stability. To verify this, we should again make sure that gauge symmetry does not offer any extra protection. To this end we note that $\Psi^k \propto \psi_1 \psi_2 ... \psi_{2k}$ is a gauge invariant quantity. Consequently, the following two terms can arise when integrating out the non-zero modes: 
\begin{equation}\label{sp(2k)}
\bar{\psi}_{1} \bar{\psi}_{2} ...\bar{\psi}_{2k}\psi_{\bar{1}} \psi_{\bar{2}} ...\psi_{\bar{2k}} +
\bar{\psi}_{\bar{1}} \bar{\psi}_{\bar{2}} ...\bar{\psi}_{\bar{2k}}\psi_{1} \psi_{2} ...\psi_{2k}, 
\end{equation}
where the first term is associated with the top edge and the second with the bottom edge of the cylinder. 
Given these terms the partition function is nonzero as can be verified explicitly. A microscopic perturbation can be constructed in a fashion analogous to the case of $SU(2N)$-based FTIs. Hence the system is not stable and there is no protected excitation within the same topological sector. In particular the bosonic Moore-Read state is shown to be unstable.

\subsection{Fermionic $M=1,\ \mathbb{Z}_k$ Read-Rezayi-based FTIs}
Let us as well analyze the parton construction of FTIs based on the $M=1$ $\mathbb{Z}_k$ Read-Rezayi states. Following Ref. \onlinecite{Wen9}, for each spin species we introduce a vector of $2k$ partons $\psi = (\psi_1,\psi_2, ... ,\psi_{2k})$ with charge $e^* = e/(k+2)$ each and additionally one charge $e\cdot k/(2+k)$ parton $\psi_0$. The electron operator is given by:
\begin{equation}
\Psi = \psi_0 \psi^T \Lambda \psi,
\end{equation}
with $\Lambda \in Sp(2k)$ as in the $M=0$ Read-Rezayi sequence.
This representation implies a $Sp(2k)\otimes U(1)$ gauge symmetry where the $U(1)$ piece 'glues' the additional $\psi_0$ parton to the rest. 

We consider the cases of $k$ odd and even separately. For the odd case the minimal flux insertion which returns the system to the same topological sector is that of $k+2$ flux quanta. Under the flux insertion the $2k$ partons with charge $e/(k+2)$ generate an even number of zero-modes per spin species, while the remaining $\psi_0$ parton of charge $e\cdot k/(k+2)$ generates $k$ of them. Thus the total number of zero-modes per spin species is odd and hence the $\mathbb{Z}_2$ topological index will be nontrivial, i.e. the system is stable. In particular we conclude that the FTI based on the $\mathbb{Z}_3$ Read-Rezayi state is stable, in agreement with the result of Ref. \onlinecite{Cappelli}.

For the $k$ being even case, the $\psi_0$ parton also generates $k$ zero modes per spin species, however now this number is even and thus the $\mathbb{Z}_2$ index is trivial and there is no protection. We can also construct the following term in the effective action, consistent with all symmetries:
\begin{gather}\nonumber \bar{\psi}_{a}\bar{\psi}_{b}\bar{\psi}_{1}\ldots\bar{\psi}_{k}\psi_{\bar{a}}\psi_{\bar{b}}\psi_{\bar{1}}\ldots\psi_{\bar{k}} + \\ \label{gapmonekeven}+\bar{\psi}_{\bar{a}}\bar{\psi}_{\bar{b}}\bar{\psi}_{\bar{1}}\ldots\bar{\psi}_{\bar{k}}\psi_{a}\psi_{b}\psi_{1}\ldots\psi_{k}, \end{gather}
where $\psi_{a,b}$ are the two zero-mode components associated with the field $\psi_0$. Hence the partition function is not zero and the $k$ even system is not topologically stable.

\subsection{3D Fractional topological insulator} 

Finally, we analyze the 3D fractional topological insulator, whose parton construction was described in Ref. \onlinecite{Senthil}. We introduce a single spinfull parton $d_{\alpha}$ of charge $e/2$ and take the free parton theory to be that of a 3D strong topological insulator\cite{FuKaneMele}. The physical operator is a spinless boson $\Phi$ of charge $e$ given by: 
\begin{equation}
\Phi = d_{\uparrow}d_{\downarrow}-d_{\downarrow} d_{\uparrow}.
\end{equation} 
The gluing is done by a local $SU(2)$ gauge field which forces spin singlets at each point in space. 

We make the assumption that the projection to spin singlets --  or equivalently the integration over the gauge field -- can be carried out at the level of the effective theory describing the surfaces of this state. We imagine taking periodic boundary conditions for the surfaces in two directions ('Corbino doughnut' geometry) and pierce two fluxes $\phi_x$ and $\phi_y$ through the holes of this thickened torus. The free parton theory of this system is by construction equivalent to that of a strong topological insulator \cite{FuKaneMele}, whose surface theory has a well known pair-switching behavior. Namely, for either $\phi_x = 0$ or $\phi_x = \pi$ the spectrum performs pair-switching\cite{FuKaneMele} as a function of $\phi_y$ and similarly so when the labels $x$ and $y$ are interchanged. Consequently, a flux insertion along $\phi_y$ for either $\phi_x = 0$ or $\phi_x = \pi$, will generate a single pair of action zero modes and therefore the free parton theory is anomalous. Finally, the projection onto singlets carried out using a fluctuating gauge field still leaves the anomaly intact -- thus this fractional topological insulator is stable.

\section{Discussion and outlook}
\label{Sec-discussion}

In this work we have analyzed the robustness of topological insulators (TIs) and fractional topological insulators (FTIs) to interactions by using field-theoretic tools, most notably anomalies and projective constructions. Topological insulators in $2D$ and $3D$ were found to be stable to the extent that they always support a low-lying excitation on each boundary, as long as the bulk gap persists and time reversal symmetry is not spontaneously broken. For fractional topological insulators obtained from parton or projective constructions, we have derived a stability criterion. It states that if the free parton theory is anomalous then the FTI is stable. Alternatively stated, the FTI is stable if the free parton spectrum performs pairswitching after the insertion of half of a full flux quantum. Here by full-flux insertion we mean the minimal flux insertion which returns the bulk to the same topological sector. This simply means that the free parton theory is in a topological insulator phase. The same criterion applies to $3D$ fractional topological insulators. Notice that our results also apply to the case of disordered systems, since the argument does not rely on the presence of translation invariance.

We have considered a variety of examples, in particular two-dimensional FTIs based on two copies of Laughlin $\nu = \frac{1}{2N+1}$ states were found to be stable as have been the ones based on fermionic $\mathbb{Z}_{2k+1}$ Read-Rezayi states with $M=1$. In contrast, the fractional topological insulators constructed based on $\nu = \frac{1}{2N}$ Lauglin states, fermionic $\mathbb{Z}_{2k}$ Read-Rezayi states with $M=1$ and bosonic $\mathbb{Z}_{k}$ Read-Rezayi states with $M=0$ were shown to be unstable. We also find that the 3D bosonic FTI constructed by projecting a regular strong topological insulator on a local singlet basis\cite{Senthil} is stable. All of these results are consistent with the previous analysis\cite{AdyLevinStability} and the very recent results of Ref. \onlinecite{Cappelli}.

While it is clear that our stability criterion is sufficient, it is less obvious that it is also necessary. In all the examples we considered, whenever the stability criterion was not fulfilled a symmetry-respecting term could be generated in the effective zero-mode action, which gapped out the zero modes. We have not proven this fact in a general case, however. It is conceivable that in certain particular cases other mechanisms could stabilize the edge.

The approach we apply offers distinct advantages. The stability criterion has a clear physical meaning, namely the presence of (at least) a single low-lying edge excitation in each topological sector. It is also easy to check for the FTIs described by parton constructions, as it can be performed at the level of a free-parton system. Conveniently, a detailed treatment of the fluctuating gauge field gluing the partons together is not necessary. Furthermore, being phrased in terms of anomalies, it is essentially a nonperturbative treatment which places TIs and FTIs within a unifying field-theoretical framework.  
 
It would be interesting to make refinements on these stability results. One direction is to consider whether the anomaly (or parton pair-switching) is indeed  also a necessary stability criterion. The Pfaffian $\nu=5/2$ state, for instance, appears trivial according to our considerations and also those of Ref. \onlinecite{Cappelli}. It is, however, unclear to us what the microscopic perturbation which trivializes this state is. More generally, one could ask about a procedure to reconstruct the appropriate perturbation based on the symmetry-respecting term in the low-energy effective action.  Also, it would be worthwhile to inquire in which cases one can ensure a true critical theory on the edge, as opposed to just a low-lying excitation. Along the same lines, it remains unclear how the robustness discussed in this work affects the bulk quasiparticles and their sensitivity to spin mixing perturbations. 

\emph{Acknowledgments}. All authors contributed equally to this work. We would like to thank Ady Stern and Erez Berg for useful discussions and reading the manuscript. The authors acknowledge the support from the Minerva Foundation. This work was supported in part by EPSRC Grant Nos. EP/I032487/1 and EP/I031014/1.

\bibliography{anomaly}

\begin{thebibliography}{10}

\bibitem{Hasan2010}
M.~Z. Hasan and C.~L. Kane,
\newblock Rev. Mod. Phys. {\bf 82}, 3045 (2010).

\bibitem{Qi2008}
X.-L. Qi, T.~L. Hughes, and S.-C. Zhang,
\newblock Phys. Rev. B {\bf 78}, 195424 (2008).

\bibitem{Liu2012}
C.-X. {Liu}, X.-L. {Qi}, and S.-C. {Zhang},
\newblock Physica E Low-Dimensional Systems and Nanostructures {\bf 44}, 906
  (2012), 1110.3420.

\bibitem{AdyLevin}
M.~Levin and A.~Stern,
\newblock Phys. Rev. Lett. {\bf 103}, 196803 (2009).

\bibitem{Maciejko}
J.~Maciejko, X.-L. Qi, A.~Karch, and S.-C. Zhang,
\newblock Phys. Rev. Lett. {\bf 105}, 246809 (2010).

\bibitem{MaciekLevinAdy}
M.~Koch-Janusz, M.~Levin, and A.~Stern,
\newblock Phys. Rev. B {\bf 88}, 115133 (2013).

\bibitem{RyuMoore}
S.~Ryu, J.~E. Moore, and A.~W.~W. Ludwig,
\newblock Phys. Rev. B {\bf 85}, 045104 (2012).

\bibitem{Wang2010}
Z.~Wang, X.-L. Qi, and S.-C. Zhang,
\newblock Phys. Rev. Lett. {\bf 105}, 256803 (2010).

\bibitem{Gurarie2011}
V.~Gurarie,
\newblock Phys. Rev. B {\bf 83}, 085426 (2011).

\bibitem{AdyLevinStability}
M.~Levin and A.~Stern,
\newblock Phys. Rev. B {\bf 86}, 115131 (2012).

\bibitem{Mudry1}
L.~Santos, T.~Neupert, S.~Ryu, C.~Chamon, and C.~Mudry,
\newblock Phys. Rev. B {\bf 84}, 165138 (2011).

\bibitem{Mudry2}
T.~Neupert, L.~Santos, S.~Ryu, C.~Chamon, and C.~Mudry,
\newblock Phys. Rev. B {\bf 84}, 165107 (2011).

\bibitem{Cappelli}
A.~{Cappelli} and E.~{Randellini},
\newblock ArXiv e-prints  (2013), arXiv:1309.2155.

\bibitem{Cappelli-misc}
During the process of writing this manuscript a new ArXiv publication
  appeared\cite{Cappelli}. Our results have an overlap with this recent work,
  although the approach itself is very different.

\bibitem{ZoharAdy2013}
Z.~Ringel and A.~Stern,
\newblock Phys. Rev. B {\bf 88}, 115307 (2013).

\bibitem{FuKane2006}
L.~Fu and C.~L. Kane,
\newblock Phys. Rev. B {\bf 74}, 195312 (2006).

\bibitem{AdyYuval}
Y.~Baum and A.~Stern,
\newblock Phys. Rev. B {\bf 85}, 121105 (2012).

\bibitem{Metlitski}
M.~A. {Metlitski}, C.~L. {Kane}, and M.~P.~A. {Fisher},
\newblock ArXiv e-prints  (2013), 1306.3286.

\bibitem{Fidkowski}
X.~{Chen}, L.~{Fidkowski}, and A.~{Vishwanath},
\newblock ArXiv e-prints  (2013), 1306.3250.

\bibitem{Potter}
C.~{Wang}, A.~C. {Potter}, and T.~{Senthil},
\newblock ArXiv e-prints  (2013), 1306.3238.

\bibitem{Senthil}
B.~Swingle, M.~Barkeshli, J.~McGreevy, and T.~Senthil,
\newblock Phys. Rev. B {\bf 83}, 195139 (2011).

\bibitem{peskin}
M.~Peskin and D.~Schroeder,
\newblock {\em An Introduction to Quantum Field Theory} (Addison-Wesley
  Publishing Company, 1995).

\bibitem{DHLee1996}
Y.-C. Kao and D.-H. Lee,
\newblock Phys. Rev. B {\bf 54}, 16903 (1996).

\bibitem{ZoharAdy}
Z.~Ringel and A.~Stern,
\newblock Phys. Rev. B {\bf 88}, 115307 (2013).

\bibitem{Fujikawa}
K.~Fujikawa,
\newblock Phys. Rev. Lett. {\bf 42}, 1195 (1979).

\bibitem{Nakahara}
M.~Nakahara,
\newblock {\em Geometry, Topology and Physics} (IOP Publishing Ltd., Bristol
  and Philadelphia, 2003).

\bibitem{Jain1989}
J.~K. Jain,
\newblock Phys. Rev. B {\bf 40}, 8079 (1989).

\bibitem{WenParton2}
B.~Blok and X.~Wen,
\newblock Nuclear Physics B {\bf 374}, 615  (1992).

\bibitem{WenParton1}
X.-G. Wen,
\newblock Phys. Rev. B {\bf 60}, 8827 (1999).

\bibitem{Wen9}
M.~Barkeshli and X.-G. Wen,
\newblock Phys. Rev. B {\bf 81}, 155302 (2010).

\bibitem{Maciejko3}
A.~Karch, J.~Maciejko, and T.~Takayanagi,
\newblock Phys. Rev. D {\bf 82}, 126003 (2010).

\bibitem{Maciejko2}
J.~Maciejko, X.-L. Qi, A.~Karch, and S.-C. Zhang,
\newblock Phys. Rev. B {\bf 86}, 235128 (2012).

\bibitem{Ye}
P.~{Ye} and X.-G. {Wen},
\newblock ArXiv e-prints  (2013), 1303.3572.

\bibitem{Adler1969}
S.~L. Adler,
\newblock Phys. Rev. {\bf 177}, 2426 (1969).

\bibitem{Thomale-misc}
It would be interesting to consider the applicability of the anomaly approach
  to related constructions in the context of spin liquids\cite{Thomale}.

\bibitem{FuKaneMele}
L.~Fu, C.~L. Kane, and E.~J. Mele,
\newblock Phys. Rev. Lett. {\bf 98}, 106803 (2007).

\bibitem{Thomale}
B.~Scharfenberger, R.~Thomale, and M.~Greiter,
\newblock Phys. Rev. B {\bf 84}, 140404 (2011).

\end{thebibliography}

\end{document}